\documentstyle[twocolumn,amssymb,aps,epsfig]{revtex}
\begin{document}

\title{\bf Absolute spin valve effect
}
\author{T.~P.~Pareek}
\address{Harisch-Chandra Research Institute, Chhatnag Road, Jhunsi,
Allahabad - 211019, India
}
\address{~
\parbox{14cm}{\rm
\medskip
We study charge transport in a two dimensional hybrid systems consisting of
nonmagnetic two dimensional electron gas with spin-orbit interaction 
sandwiched between a Ferromagnetic lead and a normal metallic 
lead (FM/2DEG/NM). 
An {\it absolute spin valve effect} is shown to exist. 
It is shown that the conductance of
such a hybrid system changes upon rotating the magnetization 
such that it always stays perpendicular to the current direction.
An ASVC ({\it Absolute Spin Valve  Coefficient })
is defined to quantify this effect and its dependence on various parameter is
studied.
}}

\maketitle

\narrowtext
In the past two decades the electronic transport properties of magnetic
layered structures have received considerable attention from the
scientific community. These structures which incorporate ultra thin films have 
shown a wealth of new effects related to the polarization of 
conduction electrons
such as {\it giant magnetoresistance} or {\it spin valve effect} \cite{dieny}. GMR or spin valve effect is observed in sandwich structures consisting of a Non-magnetic
material sandwiched between two ferromagnetic layers (F/NM/F) \cite{binasch}. 
The resistance or conductance of such a two terminal device 
depends on the relative orientation of
the magnetization of ferromagnetic layers. Hence to observe these effects one 
needs a minimum of two magnetic contact whose magnetization orientation can 
be rotated with respect to each other. 

In contrast, here we report an {\bf absolute spin valve effect}  which occurs 
with only one magnetic contact but in presence of spin-orbit interaction.
A necessary ingredient for this effect to occur is the 
simultaneous presence of {\it exchange } and {\it spin-orbit} interaction.
In particular we study a two dimensional hybrid systems shown in Fig.~1, which consist of a non-magnetic material with spin-orbit interaction 
sandwiched between 
a ferromagnetic lead and a normal non-magnetic lead (F/NM/N). 
The plane of Fig.~1 is {\it xy} plane, current is flowing along
{\it x} axis and the interface is parallel to {\it y} axis as depicted 
in Fig.~1 and the {\it z} axis is perpendicular to the plane {\it xy}.
In this natural coordinate system the magnetization direction is
given by usual spherical angle $\theta$ and $\phi$.
Our numerical simulation shows that
the resistance or conductance of this F/NM/N hybrid structure depends on the 
{\it absolute direction} of magnetization in the coordinate system shown in
fig.~1. 
Specifically we show that the conductance changes when magnetization direction
is rotated in {\it yz} plane while current is flowing along {\it x} axis. In
terms of polar coordinates it corresponds to a situation where we keep 
the azimuthal
angle $\phi$=90$^\circ$ fixed while we change the angle $\theta$.
Notice that in this geometry the current direction ({\it x} axis) always
stays perpendicular to the magnetization direction which lies in the 
{\it yz} plane.
We would like to stress that this effect is different from the usual 
spin valve effect which occurs with two magnetic contacts and corresponds to a
change in resistance as the relative angle between two magnetization direction is changed \cite{bauer}, \cite{monsma}. While in the effect discussed in present study occurs with only
one ferromagnetic contact and corresponds to a dependence of resistance
on the absolute direction of magnetization. Which we name aptly as
{\bf Absolute spin valve effect}. Also the effect discussed is different from the usual
anisotropic magnetoresistance effect which is a change in resistance when magnetization is rotated from being parallel to current direction to the perpendicular 
\cite{potter}.  For the geometry shown in Fig.~1 
this would correspond to a situation where
$\phi$=0 degree is kept fixed while $\theta$ is being changed, {\it i.e.}, 
magnetization direction rotates in {\it zx} plane. In our case magnetization always stays
perpendicular to the current.
To the best of our knowledge
the effect discussed here in this paper has not been discussed in the available
literature.

In our recent study for the 
case when both the contacts are ferromagnetic \cite{tribh1} (FM/2DEG/FM)
we pointed out the anisotropy in charge and spin transport.This effect has already been observed experimentally by young {\it et. al.} \cite{young}. 
However the 
systems considered in this paper where only one ferromagnetic 
contact is present is controversial. Infact in  Ref.\cite{larsen} and 
\cite{mole}
it was claimed that conductance for a FM/2DEG/NM systems does not depend
on the magnetization direction. However the calculation of Ref.\cite{larsen}
and \cite{mole}
was for one channel case and also the multiple reflection effect was neglected.
In contrast to these claims we show here that conductance of such a
system depends on {\it absolute direction} of magnetization, which we aptly name as {\bf Absolute Spin Valve Effect}.
Further the present study is not constrained to a particular kind of
spin-orbit interaction which was the case in Ref.\cite{larsen}
and \cite{mole}.  Also our calculation
is exact and takes the quantum effects at single particle level into account
\cite{tribh2}, \cite{tribhu3}.

The Hamiltonian of the full system sketched in Fig.~1 is,
\begin{equation}
H =\frac{\hat{p}^{2}}{2m^{*}} + V({\bf r}) +
\frac{\Delta}{2}\vec{\mu}({\bf r}) \cdot\vec{\sigma} +
H_{so}
\label{h_cont3d}
\end{equation} 
\noindent where the first two terms are usual kinetic and potential
energies while  the third and forth terms represent exchange and
spin-orbit interaction, respectively, $m^{*}$ is the effective
mass of electron, $\Delta$ the exchange splitting 
($\Delta$=0 for non-magnetic part of the structure),
$\vec{\mu}$ a
unit vector in the direction of magnetization of FMs and is given by
(cos$\phi$sin$\theta$, sin$\phi$sin$\theta$, cos$\theta$) and ${\bf {\sigma}}$ is a vector of Pauli matrices.
$H_{so}$ corresponds to the spin-orbit interaction terms present in the
middle region. For our present study we consider two type of spin-orbit 
interaction: (a) impurity induced spin-orbit interaction causing spin flips
during the momentum scattering leading to spin-relaxation mechanism know as
Elliot-Yafet spin relaxation \cite{elliot}(We will denote
this kind of interaction as Elliot-Yafet spin orbit interaction (EYSO)).   
(b) Rashba spin-orbit interaction (RSO) which arises due to asymmetry of 
confining potential \cite{rashba}.

Elliot-Yafet spin-orbit interaction arises 
due to the presence of impurities and is given by for the geometry shown in
fig.~1,
\begin{equation}
H_{so-ey}= \alpha_{ey}\sigma_{z}(p_{y}\,\partial_{x}V - p_{x}\, \partial_{y}V)
\label{elliot-yafet}
\end{equation}
\noindent where $V({\bf r})$ is potential due to impurities, 
$p_{x}$ and $p_{y}$ are momentum along $x$ and $y$ direction respectively
and $\alpha_{ey}$ is spin-orbit coupling coefficient. For strictly
two dimensional systems considered here $V({\bf r})$ depends on $x$ and $y$
coordinates only. We note here that the in strictly two dimensional case
the spin-orbit interaction given by eq.(\ref{elliot-yafet}) commutes with
$\sigma_z$ hence $z$ component of the spin is a good quantum number.

Rashba spin-orbit interaction which arises due to structural asymmetry 
and has the form,
\begin{equation}
H_{so-ra}=\frac{\alpha_{ra}}{\hbar}(\sigma_{x}p_{y} - \sigma_{y}p_{x})
\label{rashba}
\end{equation}
where $\alpha_{ra}$ is Rashba spin-orbit coupling coefficient. The strength of
Rashba spin-orbit interaction can be controlled by an externally applied gate
voltage \cite{lommer}, which led Datta and Das to propose the well know
spin transistor \cite{datta}. 

The essential difference between EYSO (eq.(\ref{elliot-yafet})) and RSO 
(eq.(\ref{rashba})) can be realized if we look at spin diffusion length
for the two cases. The spin diffusion lengths $l_{sd-ey}$ and
$l_{sd-ra}$ 
for EYSO and RSO respectively ,is given by
\begin{equation}
l_{sd-ey}=\frac{l_{el}}{\sqrt{2}\hbar\alpha_{ey}k_{f}^{2}} 
\label{lsdey}
\end{equation}
\begin{equation}
l_{sd-ra}=\frac{\hbar^{2}\pi}{2 m^{*}\alpha_{ra}}
\label{lsdra}
\end{equation}

\noindent where $l_{el}$ is elastic mean free path, $k_{f}$ is Fermi
momentum and $m^{*}$ is effective mass. 
We see that $l_{sd-ey}$ depends on the mean free path which is
due to the fact that the strength of EYSO given by (eq.(\ref{elliot-yafet}))
is determined by the impurities while $l_{sd-ra}$ is independent of
mean free path since the strength of RSO is essentially controlled by
structural asymmetry \cite{lommer}.

For numerical calculation we discretize the system sketched in fig.~1 on a square lattice of lattice constant {\it a} with $N_{x}$ sites along {\it x} axis and $N_{y}$ sites along {\it y} axis. The length and
width of the systems is $L=N_{x}a$ and $W=N_{y}a$ respectively.
Accordingly we use corresponding tight binding version of the 
Hamiltonians introduced in eq. (1), (2) and (3). In the tight binding
version kinetic energy term in eq. (1) transforms into the hopping term $t$
and the potential energy term $V({\bf r})$ and exchange coupling $\Delta$
give rise to the on-site energy term
$\epsilon_{i}$ . 
Model parameters in tight binding model are , hopping matrix element
$t\equiv(\hbar^{2}/2m^{*}a^{2})$ lattice spacing $a$ and on-site energy $\epsilon$. Impurities are modeled as Anderson disorder such that on-site
energies $\epsilon$ are distributed randomly between -V/2 and +V/2, 
where V characterizes the strength of disorder. Mean free path for two
dimensional tight binding model is given as 
$l_{el}=\frac{96\sqrt{E_{f}t}\,t a}{\pi V^{2}}$, where
$t$ and $a$ are hopping parameter and lattice spacing respectively 
\cite{tribh2},\cite{tribhu3}. The spin diffusion lengths  defined in eq.(\ref{lsdey}) and
eq.(\ref{lsdra}) can be recast in terms of tight binding model parameter
and are given as $l_{sd}^{ey}=\frac{l_{el}t}{\sqrt{2}\lambda_{ey}E_{f}}$
and $l_{sd}^{ra}=\frac{\pi\, a}{\lambda_{ra}}$, where $\lambda_{ey}=\hbar\alpha_{ey}/a^{2}$ and $\lambda_{ra}=\alpha_{ey}/2\,t\,a$ are dimensionless EYSO coupling
parameter and RSO coupling parameter respectively.
For details of tight binding form of Hamiltonian
we refer the reader to references \cite{tribh1},\cite{tribh2}, \cite{tribhu3}.

The conductance and spin resolved conductances are calculated
using Landauer-B\"uttiker \cite{but} formalism with the help of
non-equilibrium Green's function formalism\cite{tribhu3}. The two
terminal spin resolved conductance (for a given spin quantization
axis) is given by \cite{tribh2} \cite{tribhu3}
\begin{equation}
G^{\sigma \sigma'}(\epsilon_F)=
\frac{e^2}{h} Tr[\Gamma_1^{\sigma} G_{1N_{x}}^{\sigma \sigma'+}
\Gamma_{N_{x}}^{\sigma'} G_{N_{x}1}^{\sigma' \sigma -}]
\end{equation}
where $\Gamma_{1(N_{x})} $ self-energy function for the isolated
ideal leads and are given by $\Gamma_{p(q)}$=$t^{2}A_{p(q)}$,
where $A_{p(q)}$ is the spectral density in the respective lead
when it is decoupled from the structure, $G_{1N_{x}}^{\sigma
\sigma'+}$ and $G_{N_{x}1}^{\sigma' \sigma -}$ are the retarded
and advanced Green's functions of whole structure taking leads
into account. The trace is over spatial degrees of freedom. The
total conductance is sum of spin-conserved conductance and
spin-flip conductance, i.e., $G=G_{sc}+G_{sf}$ where the
spin-conserved and spin-flip conductance are $G_{sc}=G^{\uparrow
\uparrow}+G^{\downarrow \downarrow}$ and $G_{sf}=G^{\uparrow
\downarrow}+G^{\downarrow \uparrow}$ respectively.

\begin{figure}
\begin{center}
\mbox{\epsfig{file=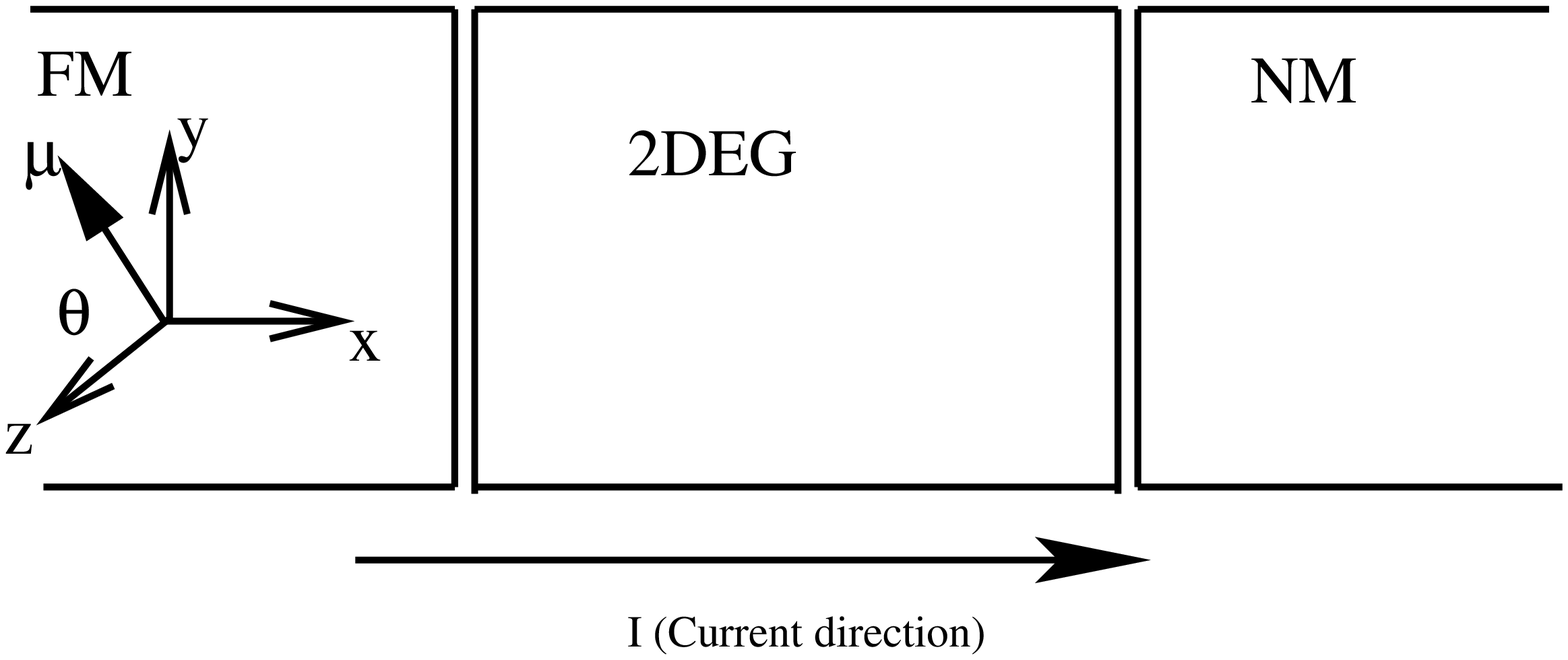,width=3in,height=1.5in,angle=0}}
\end{center}
\caption{A 2DEG connected to a Ferromagnetic and non-magnetic
ideal leads. 2DEG lies in {\it xy} plane as shown in fig. 
Magnetization direction ($\vec{\mu}$) of
ferromagnetic lead is rotated in {\it yz} plane while current is flowing
along {\it x} direction as depicted above.}
\label{Fig. 1}
\end{figure}

We first present numerical results which shows the 
{\it Absolute spin valve effect} clearly. For numerical simulation
we have taken $N_{x}=N_{y}=50$. Fermi energy  and exchange splitting 
is kept fixed at $E_{F}/t=1.0$, and $\Delta/t=0.5$,
where $t$ is the usual hopping parameter in the tight binding model.

In fig.~2 and fig.~3 total conductance ,spin conserved conductance 
and spin flip conductance are shown as a function of angle $\theta$ while
$\phi=90$ is kept fixed. 
Fig.~2 present results for RSO interaction, i.e., 
$\lambda_{ra}$=0.1, and 
$\lambda_{ey}$=0.
Fig.~3 present results for EYSO interaction, i.e., 
$\lambda_{ra}$=0, and 
$\lambda_{ey}$=0.1.
The strength of disorder potential is chosen $V/t$=1, corresponding to
a mean free path of $l_{el}=30a$.
We see that conductance changes as magnetization is rotated from {\it z} axis
to {\it y} axis. We remind the reader that we do not consider simultaneous
presence of EYSO and RSO, rather we consider the situation where either
of the two spin-orbit interactions are non zero.
\begin{figure}
\begin{center}
\mbox{\epsfig{file=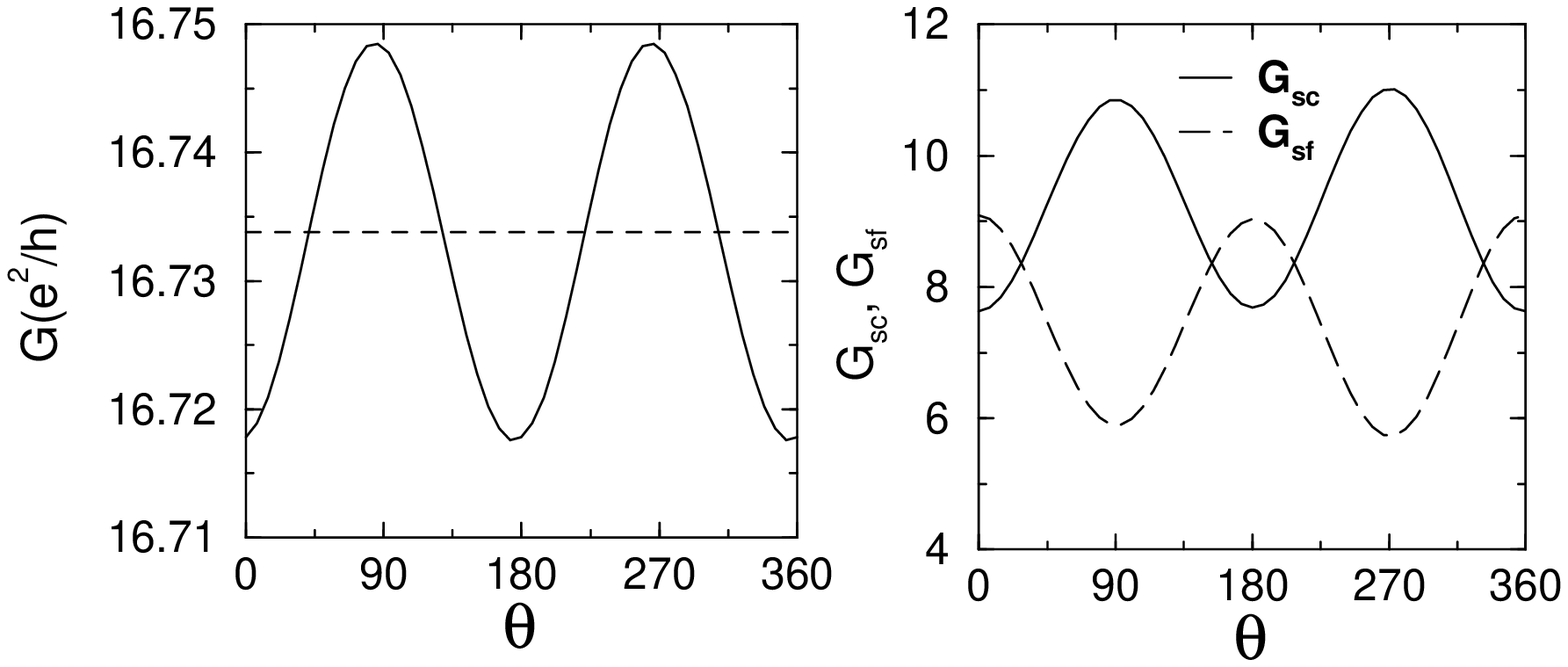,width=3in,height=2in,angle=0}}
\end{center}
\caption{Conductance, spin conserved and spin flip conductance
as a function of polar angle $\theta$ for RSO interaction.
The parameter are $\lambda_{so-ra}/t=0.1$ (spin diffusion length
$l_{sd-ra}=31\,a$), $E_{f}/t=1.0$, $V/t=1.0$ and exchange splitting
in FM is $\Delta/t=0.5$.}
\label{Fig. 2}
\end{figure}

\begin{figure}
\begin{center}
\mbox{\epsfig{file=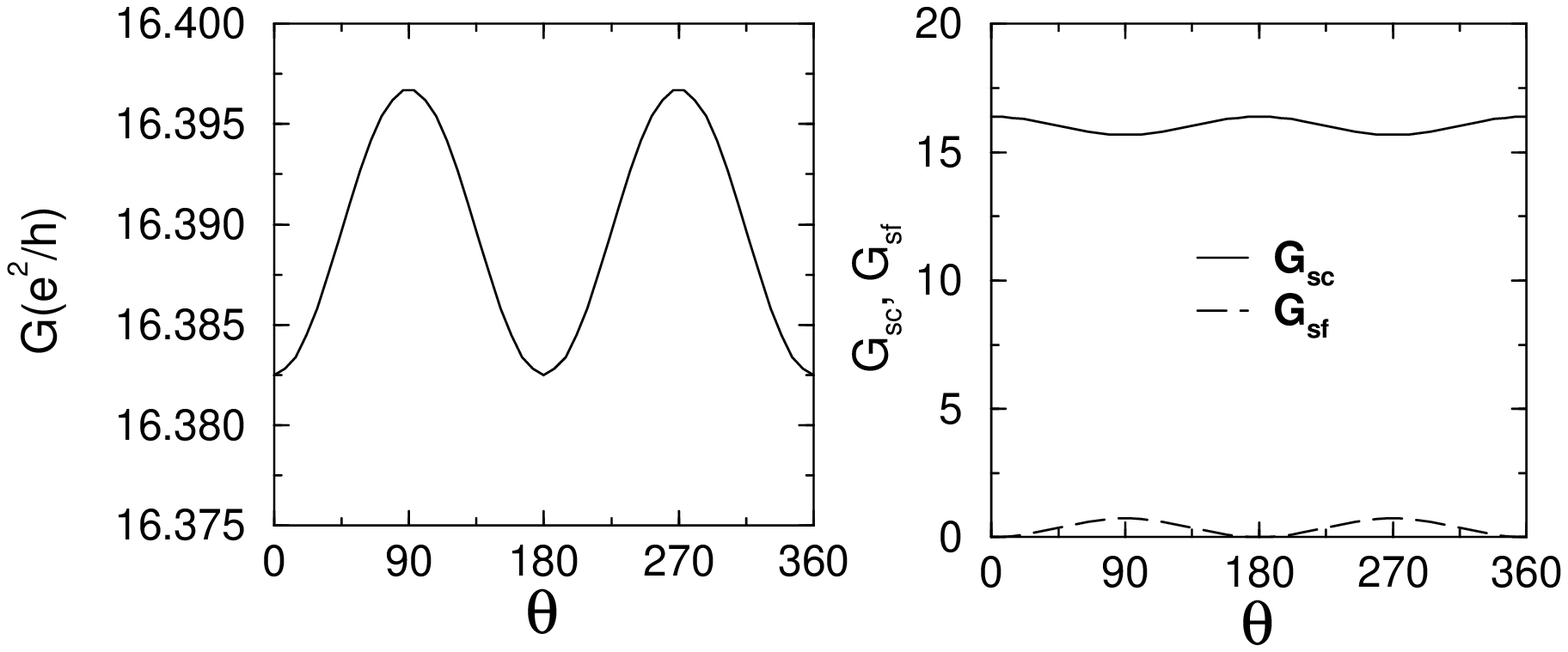,width=3in,height=2in,angle=0}}
\end{center}
\caption{Total conductance and spin conserved and spin flip conductance
as a function of polar angle $\theta$ for EYSO interaction. 
The parameter are $\lambda_{so-ey}/t=0.1$ (spin diffusion length
$l_{sd-ey}=300\,a $), $E_{f}/t=1.0$, $W/t=1.0$ and exchange splitting
in FM is $\Delta/t=0.5$.}
\label{Fig. 3}
\end{figure}

To be specific, when $\theta$=0, magnetization is parallel to 
{\it z} axis , while for $\theta=90$ 
magnetization is parallel to {\it y} axis. We would like to stress that the
current is flowing along {\it x} axis, hence the magnetization which
is being rotated in {\it yz} plane is always perpendicular to the 
current direction. Hence effect presented here is qualitatively new effect
and is different from the  usual spin valve effect and
anisotropic magnetoresistance as pointed out in
introduction. 
We see that the effect is present for both kind of spin-orbit coupling
though the order of magnitude is different. 
This can be understood if we examine the spin conserved and spin flip
conductances shown in right panels of the fig.~2 and fig.~3.
As is seen the magnitude of spin conserved and spin flip conductance
are comparable for RSO interaction (fig.~2 right panel) while for
EYSO interaction spin flip conductance is much smaller than the spin 
conserved conductances. This is so because the spin diffusion length for EYSO
is $l_{sd}^{ey}=10\,l_{el}=300\,a$ while for RSO it is 
$l_{sd}^{ra}=31\,a$, which is ten times smaller than $l_{sd}^{ey}$. 
Hence the magnitude of effect is directly determined by the spin diffusion length. 
This is also confirmed by switching off the spin-orbit
interaction, {\it i.e.}, by taking $\lambda_{ey}=\lambda_{ra}$=0,in which
case conductance shows no variation with angle $\theta$ as shown in
fig.~2(the dashed straight line in left panel).
Further we would like to point out that for EYSO the spin flip conductance
goes to zero for $\theta$=0 which is consistent with the fact that
{\it z} component of spin is conserved for EYSO in strictly 
two dimensional case.

Having demonstrated absolute spin valve effect. 
We now proceed to quantify this effect.
To this end we define, in analogy with the other magnetoresistance effect, a 
{\it Absolute Spin Valve Coefficient}
as.

\begin{equation}
ASVC= 2\frac{(G(\vec{\mu}\parallel {\bf z}) - G(\vec{\mu}\parallel {\bf y})}
{(G(\vec{\mu}\parallel {\bf z}) + G(\vec{\mu}\parallel {\bf y})}.
\label{namr}
\end{equation}

Note the ASVC coefficient defined above differs from the standard definition
in the normalization. This definition always give ASVC between 1 and -1.
The ASVC coefficient defined above measures the change in conductance 
normalized to the average conductance. Since a non zero value of ASVC requires
simultaneous presence of exchange and spin-orbit interaction, it is
natural to study ASVC as function of spin-orbit interaction. As we have 
already seen that the effect is essentially determined by spin-diffusion 
lengths
hence we plot in Fig.~4 , ASVC as a function of spin diffusion lengths which
are inversely proportional to spin-orbit coupling.
This is also motivated by the fact that RSO coupling can be externally
controlled by gate voltage \cite{lommer}.
Left panel in fig.~4 corresponds to case
where only RSO coupling is present and the
Right panel is for only EYSO interaction. The other parameter are $W/t=1$ and 
$E_{f}/t=1$ and exchange splitting in FM is $\Delta/t=0.5$. 
For Fig.~4 disorder averaging was done for 15 
different configuration. 
We see that the for weak spin-orbit interaction
ASVC shows quadratic behavior for both RSO and EYSO interaction and is 
always negative.
The magnitude of effect is of $0.1\%$ since the spin diffusion lengths are 
large compared to the systems size , which is 50 $\times$ 50.
However since the RSO coupling can be controlled by external
gate voltage and relatively large , it is desirable to see how ASVC
changes for large values of RSO coupling strength. This is shown in Fig.~5,
where we have varied RSO coupling strength $\lambda_{ra}$ over a range
such that the corresponding spin diffusion length becomes smaller than the
system size. We see the ASVC coefficient increases linearly with decreasing
spin-diffusion length and can reach values of the order of $1\%$.

In summary we have predicted an new absolute spin valve effect in 
two dimensional heterostructure with one ferromagnetic contact. The effect
exist due to simultaneous presence of exchange and spin-orbit interaction and is closely related to breaking of SU(2) symmetry in spin space due to the
presence of spin-orbit interaction. This is supported by numerical calculation where the said effect is shown to exist for two different kind of spin-orbit interaction. The ASVC is of the order of $1\%$, which is encouraging. Since 
the usual AMR of this magnitude has been experimentally measured \cite{dieny}. In light
of this we hope the predicted effect should be observable and may lead
to new spin valve devices.

\begin{figure}
\begin{center}
\mbox{\epsfig{file=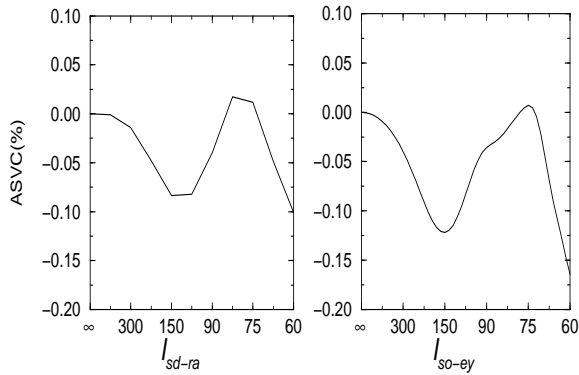,width=3in,height=2in,angle=0}}
\end{center}
\caption{ASVC coefficient (defined in eq.(\ref{namr})) as a function of
spin diffusion length. Left panel corresponds to RSO interaction
and right panel is for EYSO interaction. The other parameters are same as in fig.~2 and Fig.~3.}
\label{Fig. 4}
\end{figure}

\begin{figure}
\begin{center}
\mbox{\epsfig{file=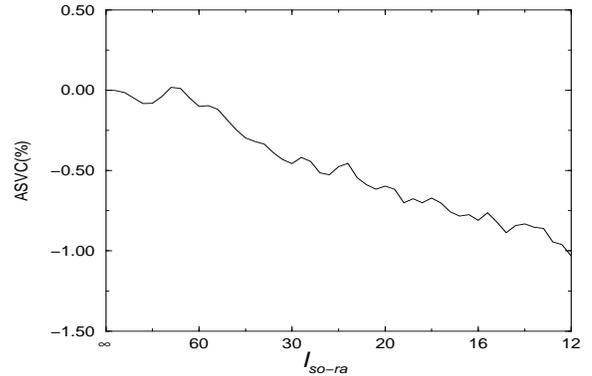,width=3in,height=2in,angle=0}}
\end{center}
\caption{ASVC coefficient (defined in eq.(\ref{namr})) as a function of
spin diffusion length for RSO interaction. The other parameters are same as 
for Fig.~4.} 
\label{Fig. 5}
\end{figure}

\end{document}